\documentclass[10pt, prd, twocolumn, showpacs, amsmath, amssymb, aps, floats, floatfix, nofootinbib]{revtex4-1}

\usepackage{bm}
\usepackage{array}
\usepackage{graphicx}
\usepackage{subfigure}
\usepackage{multirow}
\usepackage{dcolumn}
\usepackage[]{color}
\usepackage[CJKbookmarks=true,colorlinks,linkcolor=blue,anchorcolor=red,citecolor=green]{hyperref}
\usepackage{cleveref}
\usepackage{multirow}

\def\nat{Nature\ }
\def\aap{Astron.\ Astrophys.\ }
\def\apj{Astrophys.\ J.\ }
\def\apjl{Astrophys.\ J.\ Lett.\ }
\def\apjs{Astrophys.\ J.\ Supp.\ }
\def\aj{Astron.\ J.\ }
\def\mnras{Mon.\ Not.\ Roy.\ Astron.\ Soc.\ }
\def\physrep{Phys.\ Rept.\ }
\def\prd{Phys.\ Rev.\ D\ }

\def\araa{Annu.\ Rev.\ Astron.\ Astrophys.\ }
\def\jcap{J.\ Cosmol.\ Astropart.\ Phys.\ }

\def\km{\mathrm{km}}

\def\GeV{\mathrm{GeV}}

\def\cm{\mathrm{cm}}
\def\s{\mathrm{s}}
\def\sr{\mathrm{sr}}
\newcolumntype{p}{D{,}{\pm}{-1}}

\begin{document}
\title{Constraint on the velocity dependent dark matter annihilation cross section
from Fermi-LAT observations of dwarf galaxies}

\author{Yi Zhao$^{1,2}$}
\author{Xiao-Jun Bi$^{2}$}
\author{Huan-Yu Jia$^{1}$}
\author{Peng-Fei Yin$^{2}$}
\author{Feng-Rong Zhu$^{1}$}

\address{
$^1$ Southwest Jiaotong University, Chengdu 610031, China\\
$^2$ Key Laboratory of Particle Astrophysics,
Institute of High Energy Physics, Chinese Academy of Sciences,
Beijing 100049, China\\}

\begin{abstract}
The $\gamma$-ray observation of dwarf spheroidal satellites (dSph's)
is an ideal approach for probing the dark matter (DM) annihilation
signature. The latest Fermi-LAT dSph searches have set stringent
constraints on the velocity independent annihilation cross section
in the small DM mass range, which gives very strong constraints on the
scenario to explain the AMS-02 positron excess by DM annihilation.
However, the dSph constraints would change in the velocity dependent
annihilation scenarios, because the velocity dispersion in the dSph's
varies from that in the Milky Way. In this work, we use a likelihood
map method to set constraints on the velocity dependent annihilation
cross section from the Fermi-LAT observation of six dSph's. We consider
three typical forms of the annihilation cross section, {\it i.e.}
p-wave annihilation, Sommerfeld enhancement, and Breit-Wigner resonance.
For the p-wave annihilation and Sommerfeld enhancement, the dSph limits
would become much weaker and stronger compared with those for the
velocity independent annihilation, respectively. For the Breit-Wigner
annihilation, the dSph limits would vary depending on the model parameters.
We show that the scenario to explain the AMS-02 positron excess by
DM annihilation is still viable in the velocity dependent cases.
\end{abstract}



\maketitle

\section{Introduction}
Numerous astrophysical and cosmological observations have shown that the Universe is composed of approximately $4.8\%$ baryons, $25.8\%$ cold dark matter (DM)
and $69.3\%$ dark energy \cite{2015arXiv150201582P}. A kind of popular candidate for cold DM is the so-called weakly interacting massive particle (WIMP)\cite{1996PhR...267..195J,2000RPPh...63..793B,2005PhR...405..279B}. In the thermal freeze-out scenario, the current abundance of WIMPs can explain
the observed DM relic density \cite{2012PhRvD..86b3506S}. Today WIMP annihilations can directly produce $\gamma$ rays, or indirectly produce
$\gamma$ rays through the cascade decay, final state radiation, and inverse Compton scattering processes. These $\gamma$ rays
should be dominantly generated in the areas with high DM densities, and then can be captured
by terrestrial and satellite experiments, such as the space-borne $\gamma$-ray detector Fermi Large Area Telescope (Fermi-LAT) \cite{2009ApJ...697.1071A}.

Many studies have been performed to investigate the $\gamma$-ray emission from DM annihilations in several astrophysical sources, such as the galactic halo \cite{2011PhRvD..84l3005H,2012ApJ...761...91A,2012PhRvD..86h3511A,2010PhRvL.104i1302A,
2012PhRvD..86b2002A,2012JCAP...08..007W,2013PhRvD..88h2002A}, galaxy clusters
\cite{2010JCAP...05..025A}, and galactic DM substructures \cite{2012A&A...538A..93Z,2012ApJ...747..121A,
2012JCAP...11..050Z}.
Among these sources, the nearby dwarf spheroidal satellites (dSph's) are ideal for probing the DM annihilation $\gamma$-ray signatures, because of their high DM densities and lack of conventional astrophysical $\gamma$-ray sources \cite{1998ARA&A..36..435M,2009ApJ...696..385G}. As no significant $\gamma$-ray
excess has been found in the Fermi-LAT dSph observations, the stringent upper limits on the DM annihilation cross section have been set in the literature \cite{2010ApJ...712..147A,2011PhRvL.107x1302A,2011PhRvL.107x1303G,2012PhRvD..86b3528C,
2012PhRvD..86b1302G,2012APh....37...26M,2012PhRvD..86f3521B,2012JCAP...11..048H,2014PhRvD..89d2001A,
2015arXiv150302641F,2013JCAP...03..018S}\footnote{DSph constraints on the lifetime of decaying DM can be found in Ref. \cite{2010JCAP...12..015D,2015arXiv151000389B}.}.

For the ordinary DM s-wave annihilation process which is independent of the DM relative velocity, the observed DM relic density can be achieved with a thermally averaged cross section of $\langle \sigma v \rangle \sim 3\times 10^{-26}$cm$^3$s$^{-1}$. However, the s-wave annihilation cross section has been stringently constrained by the Fermi-LAT $\gamma$-ray searches from dSph, and should be smaller than the ``canonical" value $\sim 3\times 10^{-26}$cm$^3$s$^{-1}$ for the DM mass range of $\sim \mathcal{O}(1)-\mathcal{O}(10)$ GeV for the DM annihilations to $b\bar{b}$ or $\tau^+\tau^-$ \cite{2014PhRvD..89d2001A}. Generally, $\langle \sigma v \rangle$ depends on the DM relative velocity. For example, $\langle \sigma v \rangle$ can be expanded to a form of $a+b\langle v^2 \rangle+ \mathcal{O}(v^4)$ in the nonrelativistic limit. Only when the s-wave annihilation is dominant, $\langle \sigma v \rangle$ is a constant.
If the p-wave annihilation is not negligible, $\langle \sigma v \rangle$ would be suppressed at small DM velocities. In this case, $\langle \sigma v \rangle$ in dSph's would be smaller than those in the local halo or in the early Universe as a result of a small DM velocity dispersion of $\sim \mathcal{O}(1)-\mathcal{O}(10)$ km s$^{-1}$ in dSph's \cite{2009ApJ...704.1274W}. Consequently, the constraints for velocity suppressed DM annihilations from dSph $\gamma$ ray searches is much weaker.

An interesting hint of the DM signature is from the cosmic-ray electron/positron observations.
If the anomalous electrons and positrons observed by PAMELA \cite{2009Natur.458..607A} and AMS02 \cite{2013PhRvL.110n1102A} are produced by DM annihilation, the local DM annihilation cross section in the Galaxy should be of the order $\sim 10^{-24}-10^{-23}$cm$^3$s$^{-1}$, which is approximately 2 or 3 orders of magnitude larger than the canonical value in the early Universe. This discrepancy can be explained in the velocity dependent annihilation scenarios. For example, an exchange of a new light boson between two initial heavy DM particles may confer an additional factor of $1/v$ or $1/v^2$ to the annihilation cross section. This effect is the so-called Sommerfeld enhancement \cite{1931AdPh.403.257}, and can be used to simultaneously explain the anomalous cosmic-ray electrons/positrons observed by the PAMELA and the DM relic density \cite{2005PhRvD..71f3528H,2007PhLB..646...34H,
2007NuPhB.787..152C,2009NuPhB.813....1C,2009PhRvD..79a5014A,2009PhLB..671..391P,2009PhRvD..79h3523L,2010PhLB..687..275D,
2010PhRvD..82h3525F,2010JPhG...37j5009C,2010JCAP...02..028S}. In this scenario, the dSph searches would set strong constraints on the annihilation cross section as result of the small velocity in dSph's \cite{2012PhRvD..86h3534S,2010PhRvD..82l3503E,2009JCAP...12..011Y} (for other cosmological limits, see e.g. \cite{2009PhRvD..80d3526S,2010JCAP...10..023Y,2011PhRvD..83l3511H}). Another important scenario for the velocity dependent annihilation is the "Breit-Weigner" scenario \cite{2009PhRvD..79f3509F,2009PhRvD..79i5009I,2009PhRvD..79e5012G,2009PhLB..678..168B}. In this scenario, the DM particles annihilate via a pole, which lies near twice the DM mass. Depending on the model parameters, the annihilation cross section can also be significantly enhanced at lower velocities.

In this work, we study the constraints on the velocity dependent DM annihilation cross sections from the Fermi-LAT dSph observations. A simple approach is to require that the DM signature should not exceed the observed $\gamma$ ray flux upper limit. However, the derived constraint without the spectral information would de very conservative. In Ref. \cite{2013JCAP...03..018S}, the authors proposed a convenient and
flexible likelihood map method to derive limits from the Fermi-LAT dSph's observations for any given shape of the initial DM-induced $\gamma$-ray spectrum. In this work, we adopt a similar method to construct the likelihood maps for six dSph's with large $J-$factors. Then we can easily obtain the total likelihood via these likelihood maps, and set constraints on the DM annihilation cross section.

This paper is organized as follows.
In Sec. II, we give a
detailed description of the likelihood map method. In Sec. III, we consider three typical forms of the velocity dependent DM annihilations, and use the likelihood map method to set constraints. Section IV is the summary.

\section{Fermi-LAT data analysis}
A maximum likelihood method has been developed for the $\gamma$-ray source analysis according to
the limited photon statistics and the dependence of Fermi-LAT performance
on the incident photon angle and energy.

For every dSph, we divide Fermi-LAT observational data into several energy bins. In the likelihood calculation, the $\gamma$-ray flux from DM annihilation in each bin is assumed to be a constant. This ``energy-flux-likelihood" cube forms the likelihood map for a certain dSph. Through the use of this likelihood map, the likelihood for any $\gamma$-ray spectrum shape can be easily obtained in the whole energy range.

The expected $\gamma$-ray signature flux from dSph DM annihilation
can be expressed as
\begin{eqnarray}\label{eq:zy2}
\phi(E) = \frac{\left<\sigma v\right>}{8\pi m_{DM}^2}\times\frac{dN_\gamma}{dE_\gamma}\times J ,
\end{eqnarray}
where $\left<\sigma v\right>$ is the thermally averaged annihilation cross section,
$m_{DM}$ is the DM mass, $\frac{dN_\gamma}{dE_\gamma}$ is the differential
$\gamma$-ray spectrum in one DM pair annihilation, and the $J$ factor is
the line-of-sight integration for the DM distribution,
i.e. $J=\int \rho^{2}(l)dld\Omega$.
Note that $\frac{dN_\gamma}{dE_\gamma}$ should be a sum of the photons from all possible DM
annihilation final states according to the DM model. Here we only consider the $\gamma$-ray contribution from a certain annihilation channel through the use of PPPC \cite{2011JCAP...03..051C,2011JCAP...03..019C}.
Then we define a variable $C_i=\frac{dN_\gamma}{dE_\gamma}|_{\sqrt{E_{i}E_{i+1}}}$ in each energy bin, where $i$ is the energy bin index.

The $J$ factor can be derived from the observed line-of-sight velocities of the stars through the use of the Jeans equation \cite{2004PhRvD..69l3501E,2008ApJ...678..614S,
2009JCAP...06..014M}. Numerous works \cite{2009JCAP...06..014M,2013PhR...531....1S}
have indicated that the $J$ factor is not sensitive to the certain DM density profile. For instance,
the $J$ factor of Draco for Navarro-Frenk-White is similar with that for the Burkert profile.
We take the values of $J$ factors and their uncertainties from Table I of Ref.\cite{2014PhRvD..89d2001A}, which are provided
by the results from Refs. \cite{2015MNRAS.451.2524M,2007ApJ...670..313S,2011ApJ...733...46S,
2005ApJ...631L.137M,2011AJ....142..128W}.

For a certain set of $m_{DM}$, $\left<\sigma v\right>$ and $J$ factor inputs, the combined likelihood in all energy bins for the $j$th dSph is estimated as
equation
\begin{eqnarray}\label{eq:zy3}
L_{j}&&=\prod_{i}L_{i}(\phi_{i}|D) \nonumber\\
&&\times \frac{1}{\ln(10)J_{obs,j}\sqrt{2\pi}\sigma_{j}}e^{-[\log_{10}(J_j)-\log_{10}(J_{obs,j})]^{2}/2\sigma_{j}^{2}},
\end{eqnarray}
where $i$ denotes the $i$th energy bin, $\phi_i$ is the DM signature flux, $J_{obs,j}$ is the calculated $J$ factor with a error of $\sigma_j$.
 For a given $\left<\sigma v\right>$ and $m_{DM}$, the $J_j$ is chosen to make the $L_j$ reach a maximum. Then we can get a
``cross section-likelihood" table for the $j$th dSph. The combined
likelihood of all dSph's can be calculated as
\begin{eqnarray}\label{eq:zy4}
L=\prod_{j}L_{j}.
\end{eqnarray}
Through the use of this combined cross section-likelihood table, we can set $95\%$ confidence level (C.L.) upper limit
on energy flux and find the value of $\left<\sigma v\right>$ by requiring that the corresponding log-likelihood has decreased by $2.71/2$ from its maximum \cite{1953Biometrika.40.306,2005NIMPA.551..493R}.

\begin{figure*}[!htb]
\centering
\includegraphics[width=1.0\columnwidth, angle=0]{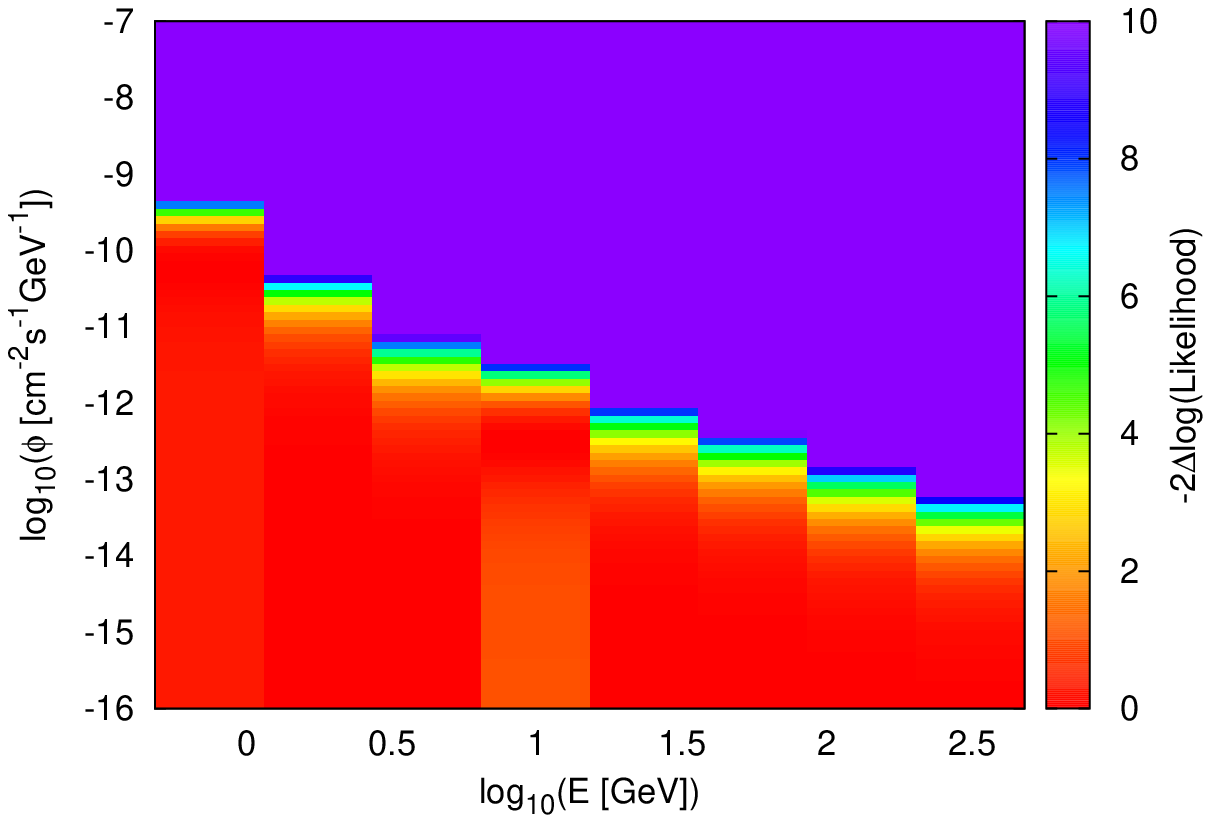}
\includegraphics[width=1.0\columnwidth, angle=0]{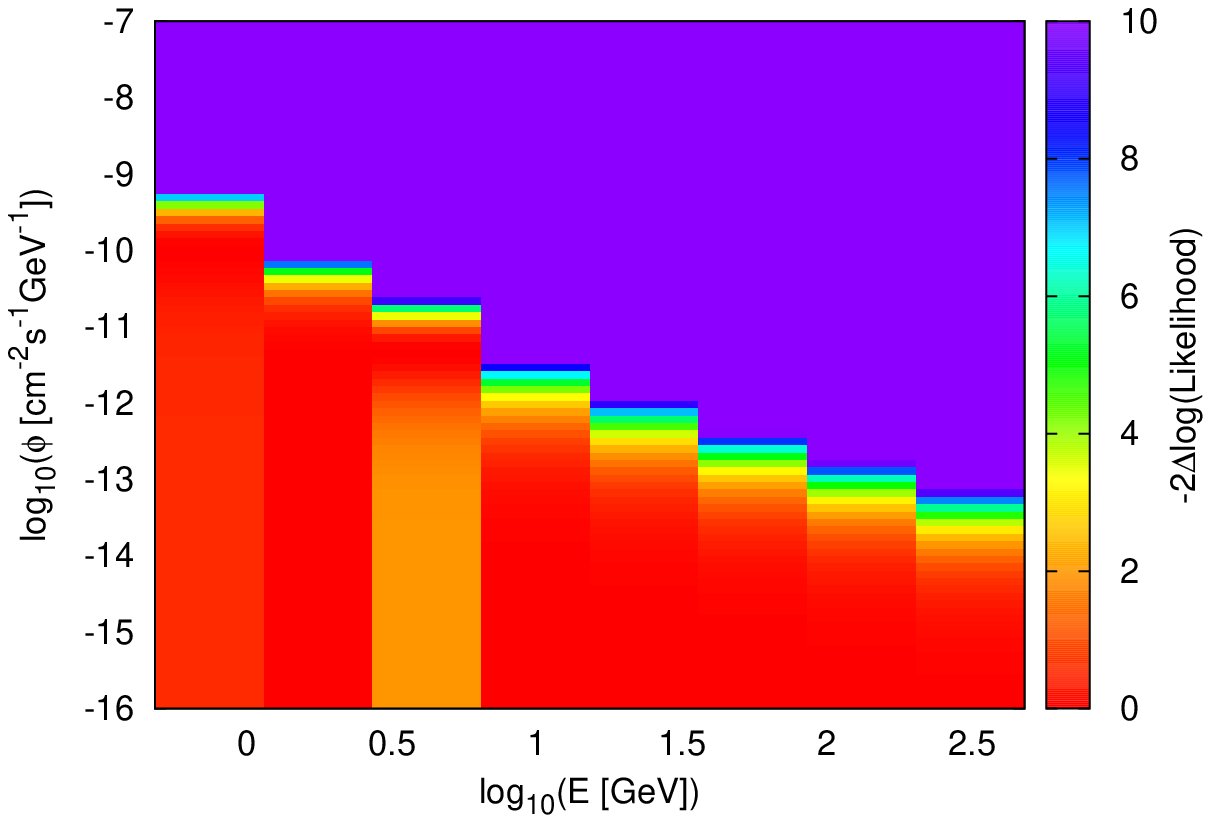}
\caption{The likelihood maps of Draco (left) and Segue1 (right).}
\label{fig:zy1}
\end{figure*}

In Ref.\cite{2013JCAP...03..018S}, the authors combined the likelihood of all selected dSph in the $i$th energy bin, and then constructed the combined likelihood map in the whole energy range. Hence the dSph uncertainties have been repeatedly included in each energy bin. The correlations between different energy bins are also not taken into account. Contrary to that approach, we first combine the likelihood of all energy bins $L_i$ for
the $j$th dSph, and then construct a total likelihood. For example, we show the likelihood maps of the Draco and Segue 1 in Fig \ref{fig:zy1}. The grid scan of the likelihood map is preformed in eight logarithmical energy bins in the range of $0.5-500$ GeV, and 300 logarithmical bins in the range of $10^{-30}-10^{-1}$cm$^{-2}$s$^{-1}$GeV$^{-1}$. The color bar denotes
the value of $-2\Delta\log L$ for the given $m_{DM}$ and signature flux. The sensitivity
of a combined dSph analysis is dominantly determined by the dSph's with large $J$ factors.
In this analysis, we consider six dSph's, i.e. the Draco, Segue 1, Coma Berenices, Willman 1, Ursa Major II and Ursa Minor, because the sensitivity
of a combined dSph analysis would be dominantly determined by such dSph's with large $J$ factors \cite{2014PhRvD..89d2001A}. The mean values and
uncertainties of the $J$ factors of these six dSph's are listed in Table \ref{tab:zy1}.
\begin{table}[!htb]
\belowcaptionskip=1em
\centering
\caption{Measured $J$ factor with error.}
\setlength\tabcolsep{1.0em}
 \begin{tabular}{lcc}
  \hline\hline\noalign{\smallskip}
   \multirow{2}{*}{Name}      & $\log_{10}(J^{\mathrm{NFW}}_{obs})$  & \multirow{2}{*}{Reference} \\
             &$(\log_{10}[\GeV^{2}\cm^{-5}\sr])$   \\
  \hline\noalign{\smallskip}
  Draco           & $18.8\pm0.16$    &\cite{2005ApJ...631L.137M}  \\
  Segue 1         & $19.5\pm0.29$    &\cite{2011ApJ...733...46S}  \\
  Coma Berenices  & $19.0\pm0.25$    &\cite{2007ApJ...670..313S}  \\
  Willman 1       & $19.1\pm0.31$    &\cite{2011AJ....142..128W}  \\
  Ursa Major II   & $19.3\pm0.28$    &\cite{2007ApJ...670..313S}  \\
  Ursa Minor      & $18.8\pm0.19$    &\cite{2005ApJ...631L.137M}  \\
  \noalign{\smallskip}\hline\hline
\end{tabular}
\label{tab:zy1}
\end{table}

Our work is based on the latest published SCIENCE TOOLS version v10r0p5. We employ six-year Fermi-Lat data recorded from
$2008-08-04$ to $2014-08-05$ with the Pass8 photon data selection in the analysis. The events from the Pass8 SOURCE-class
in the energy band between $500$ MeV and $500$ GeV are adopted. In order to reduce the $\gamma$-ray
contamination from the earth limb, the events with zenith angles larger than $100^{\circ}$ are rejected,
and the recommended filter cut $(DATA\_QUAL>0, LAT\_CONFIG==1)$ are applied. Each dSph is taken as a point source. We create $10^{\circ} \times 10^{\circ}$ square region as the region of interest around the
dSph center into $0.1^{\circ}$ pixels and eight logarithmical bins of energy from
$500$ MeV to $500$ GeV. We use the galactic $\gamma$-ray diffuse model gll\_iem\_v06.fit and the
isotropic extragalactic $\gamma$-ray diffuse spectrum iso\_P8R2\_SOURCE\_V6\_v06.txt as the diffuse background. The third LAT source catalog (3FGL) \cite{2015ApJS..218...23A} is taken to deal with the point $\gamma$-ray sources.
In the analysis, we carry out a global fit over the entire energy range and then fix all the parameters except the normalization of two diffuse backgrounds in each energy bin.
The instrument response functions P8R2\_SOURCE\_V6 have been set corresponding to the above
LAT data selection.

\section{limits on velocity dependent annihilation cross sections}

In the previous works, the DM annihilation cross section is assumed to be a velocity independent constant.
The primary goal of this work is exploration of the $\gamma$-ray observation limits from dSph on the velocity dependent annihilation cross section. We take the DM annihilation cross section as some typical velocity dependent forms, such as those in the p-wave annihilation, Breit-Wigner and Sommerfeld scenarios.

The thermally averaged annihilation cross section in the nonrelativistic limit can be calculated as
\begin{eqnarray}\label{eq:zy5}
\left<\sigma v\right>&=&\int f(\overrightarrow{v_1})f(\overrightarrow{v_2})(\sigma v_{rel})d^3\overrightarrow{v_1}d^3\overrightarrow{v_2} \nonumber\\
&=&\int \sqrt{\frac{2}{\pi}} \frac{1}{v_{p}^3} v_{rel}^2 e^{-\frac{v_{rel}^2}{2v_{p}^2}}(\sigma v_{rel})dv_{rel},
\end{eqnarray}
where $f(\overrightarrow v)$ is the DM velocity distribution, which is
assumed as the Maxwell-Boltzmann form.
$v_{rel}=|\overrightarrow{v_1}-\overrightarrow{v_2}|$ is the relative velocity of two initial DM particles, and
$v_p$ is the most probable velocity. The relation of $v_p$ and the line-of-sight velocity dispersion $v_0$ can be approximately expressed by
\begin{eqnarray}\label{eq:zy8}
v_p\approx \sqrt{2}v_0.
\end{eqnarray}
The velocity dispersions of the six dSph's considered in this work are listed in Table \ref{tab:zy2} \cite{2009ApJ...704.1274W}.
\begin{table}[!htb]
\belowcaptionskip=1em
\centering
\caption{dSph velocity dispersions.}
\setlength\tabcolsep{2.0em}
 \begin{tabular}{lcc}
  \hline\hline\noalign{\smallskip}
   \multirow{2}{*}{Name}      & $v_0$  & \multirow{2}{*}{Reference} \\
             &$(\km\s^{-1})$   \\
  \hline\noalign{\smallskip}
  Draco           & $9.1\pm1.2$    &\cite{2008ApJ...684.1075M,2007ApJ...667L..53W}  \\
  Segue 1         & $4.3\pm1.2$    &\cite{2008ApJ...684.1075M,2009ApJ...692.1464G}  \\
  Coma Berenices  & $4.6\pm0.8$    &\cite{2008ApJ...684.1075M,2007ApJ...670..313S}  \\
  Willman 1       & $4.3\pm1.8$    &\cite{2008ApJ...684.1075M,2007MNRAS.380..281M}  \\
  Ursa Major II   & $6.7\pm1.4$    &\cite{2008ApJ...684.1075M,2007ApJ...670..313S}  \\
  Ursa Minor      & $9.5\pm1.2$    &\cite{1995MNRAS.277.1354I,2009ApJ...704.1274W}  \\
  \noalign{\smallskip}\hline\hline
\end{tabular}
\label{tab:zy2}
\end{table}

Note that the DM velocity distribution is assumed to be an isotropic Maxwell-Boltzmann distribution in Eq. \ref{eq:zy5}. For a given DM density profile and an isotropic velocity distribution, the most probable velocity and one-dimensional DM velocity dispersion can be calculated from the Jeans equation \cite{2010PhRvD..82l3503E}. In principle, a precise DM velocity distribution can be derived from N-body simulations. Many analyses based on high resolution DM only simulations have shown that the DM velocity distributions of Milky Way-like halos are anisotropic and deviate from the standard Maxwell-Boltzmann distribution \cite{2006JCAP...01..014H,2009MNRAS.395..797V,2010JCAP...02..030K,2013ApJ...764...35M}. In recent years, more realistic DM velocity distributions have been extracted from simulations including the baryonic effects during the galaxy formation process \cite{2010JCAP...02..012L,2014ApJ...784..161P,2015arXiv150304814B,2016arXiv160104707B}. These studies also confirmed that the DM velocity distribution is not a standard Maxwell-Boltzmann distribution. In order to determine the precise DM velocity distributions of the Galaxy and dSph's, further studies are required. Here we take the standard Maxwell-Boltzmann distribution as an approximation.

Unlike the velocity independent case, where $\left<\sigma v\right>$
is a universal value, now $\left<\sigma v\right>$ depends on the velocity
of the DM particles. This means
DM particles may have different $\left<\sigma v\right>$ value in each dSph.
Thus the strategy of setting constraints on
the DM annihilation cross section is model
dependent. For example, if we can expand $\left<\sigma v\right>$
in the form like $\left<\sigma v\right> = a + b v^2$, we get constraints
on the two constants of $a, ~b$ in each dSph. More severe constraints on
$a, ~b$ can be obtained by combining the constraints from several dSph's.
Furthermore the constraints on the local DM annihilation cross section
$\left<\sigma v\right>$ can be
obtained by taking the velocity dispersion in the Milky Way.
In the following we present our constraints on $\left<\sigma v\right>$
for three typical velocity dependent forms.
The results have been translated to the local $\left<\sigma v\right>$
in the Solar system.

Note that the dSph $J$ factor is also determined by the
density profile which is derived by fitting to the observed velocity
distribution of the luminous matter. Therefore, the velocity dependent
factor in the dSph DM annihilation cross section is correlated to
the $J$ factor (see e.g. \cite{2010PhRvD..82i5007C,2011PhRvD..84g5004C}). A detailed analysis of this
correlation is complex and will be left to future study.

\subsection{p-wave annihilation}
In the WIMP scenario, the DM annihilation cross section can be generically expanded by $v^2_{rel}$ in the nonrelativistic limit. We parameter the annihilation cross section as
\begin{eqnarray}\label{eq:zy9}
\sigma v_{rel}= a+bv_{rel}^2,
\end{eqnarray}
where $a$ is the dominant contribution from the s-wave annihilation process and $b$ is the contribution from the p-wave annihilation process. The higher-order contributions depending on $\mathcal{O} (v^{2n}_{rel})$ with $n>1$ have been neglected here.

\begin{figure*}[!htb]
\centering
\includegraphics[width=0.9\columnwidth, angle=0]{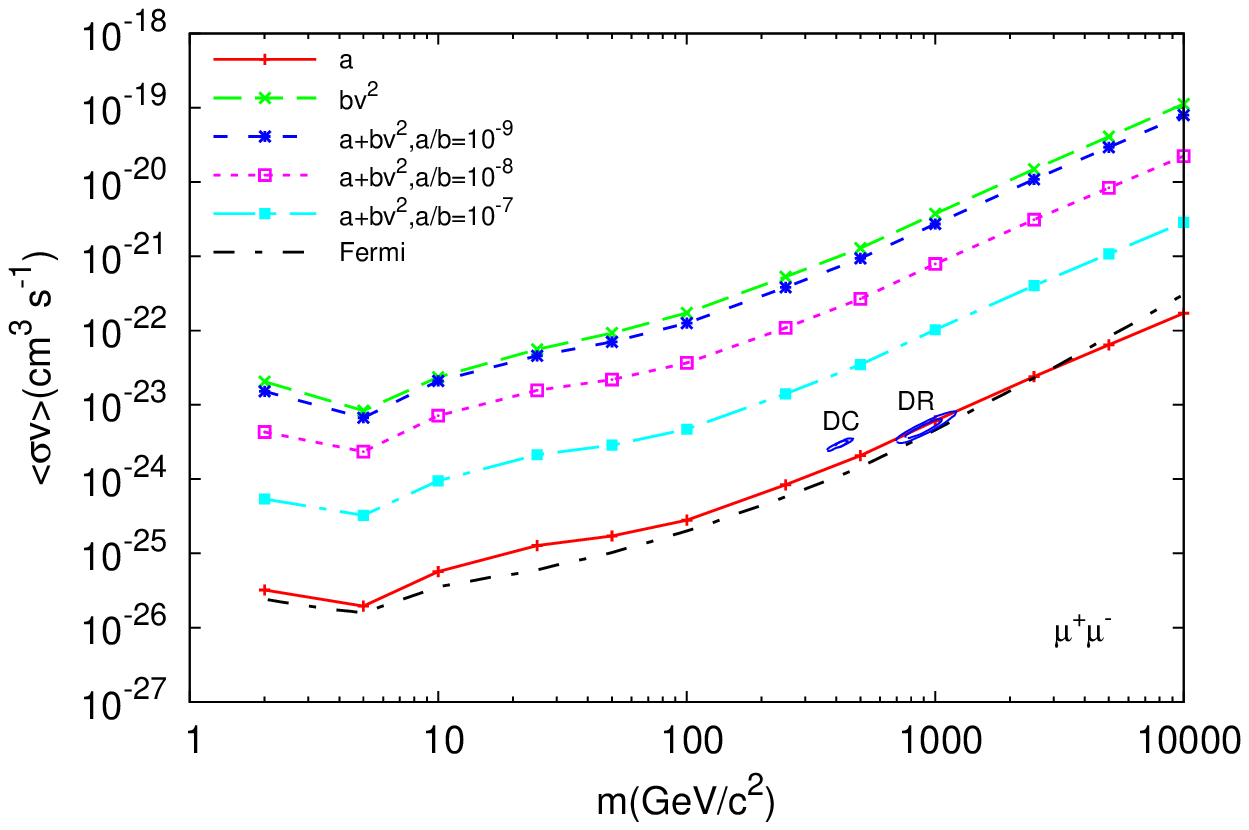}
\includegraphics[width=0.9\columnwidth, angle=0]{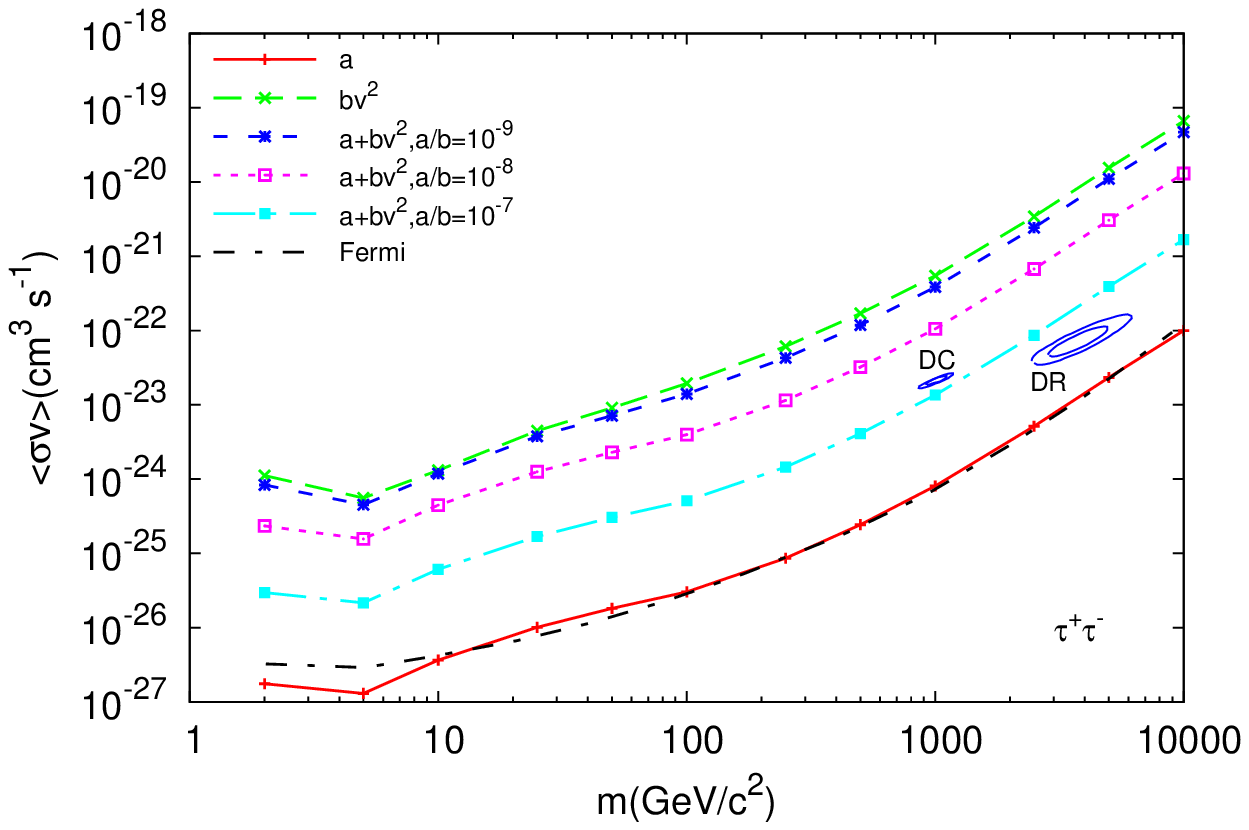}
\includegraphics[width=0.9\columnwidth, angle=0]{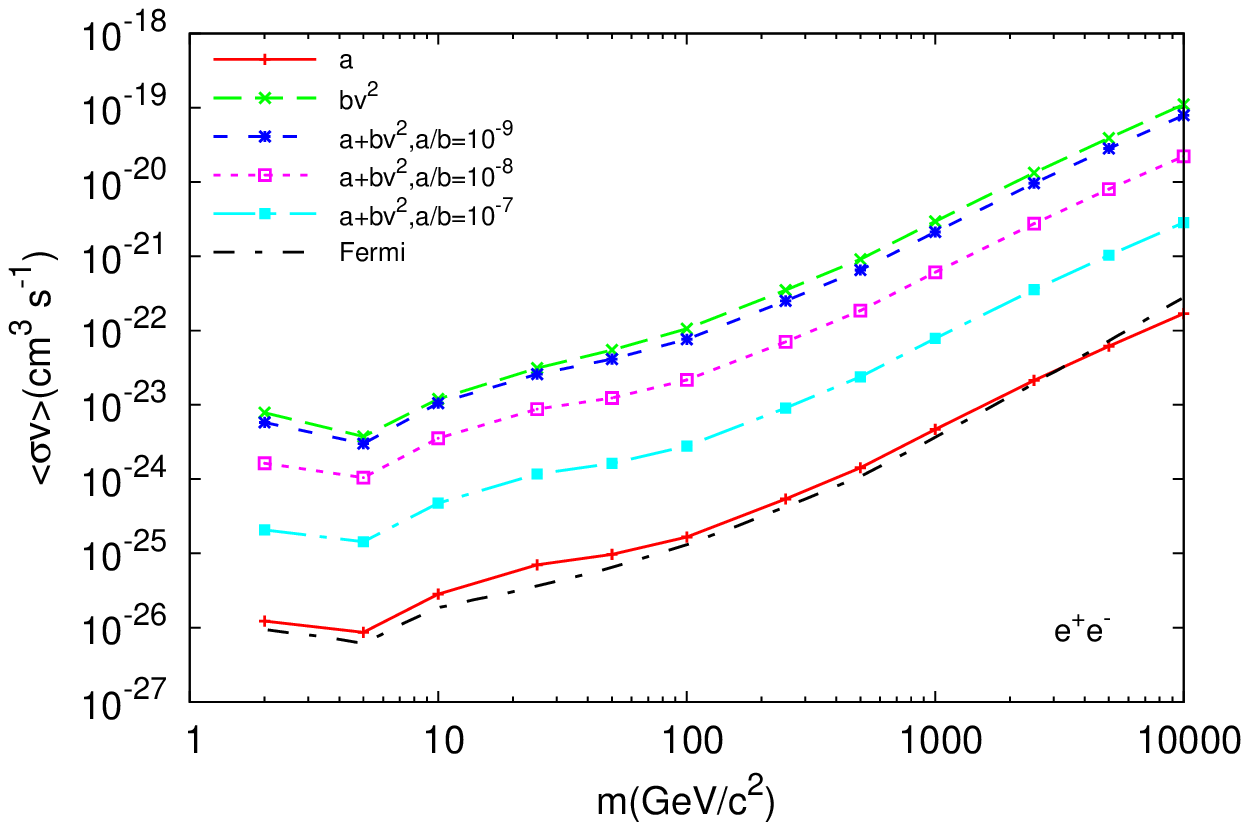}
\includegraphics[width=0.9\columnwidth, angle=0]{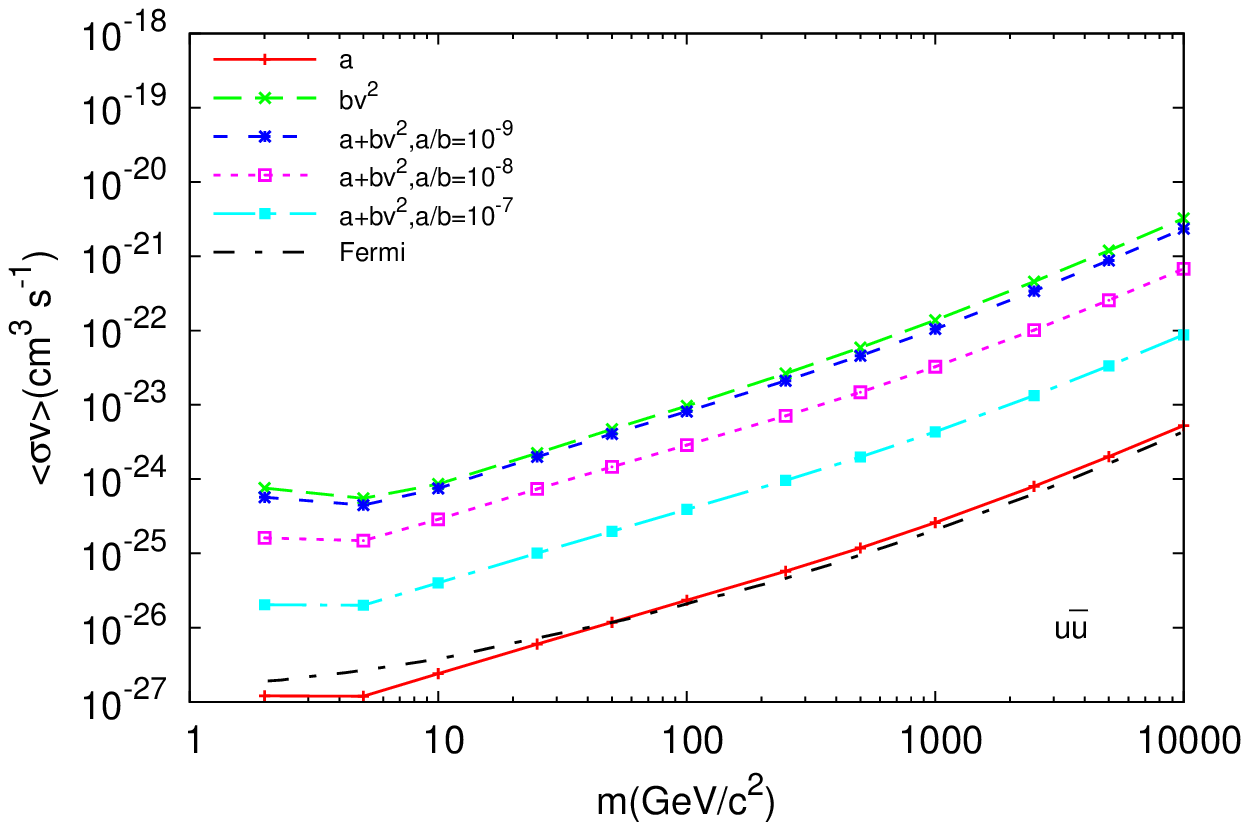}
\includegraphics[width=0.9\columnwidth, angle=0]{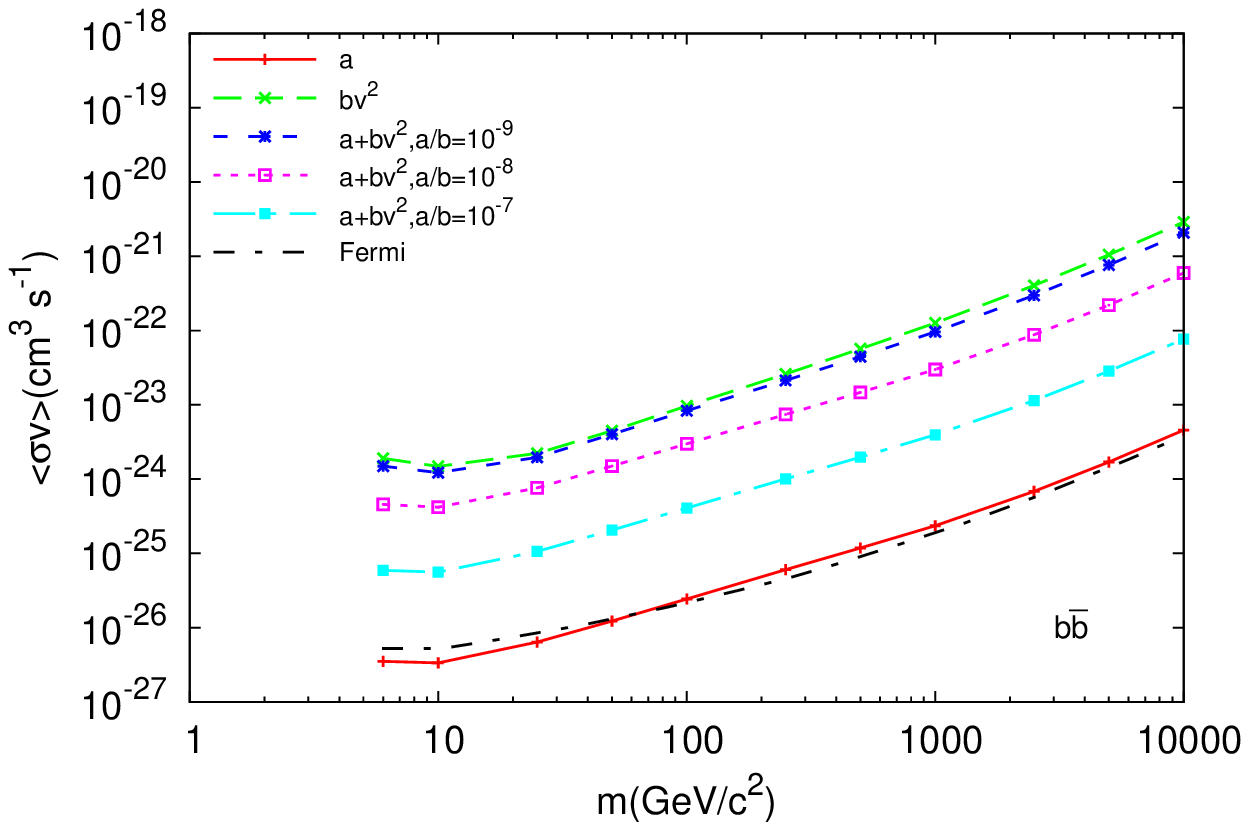}
\includegraphics[width=0.9\columnwidth, angle=0]{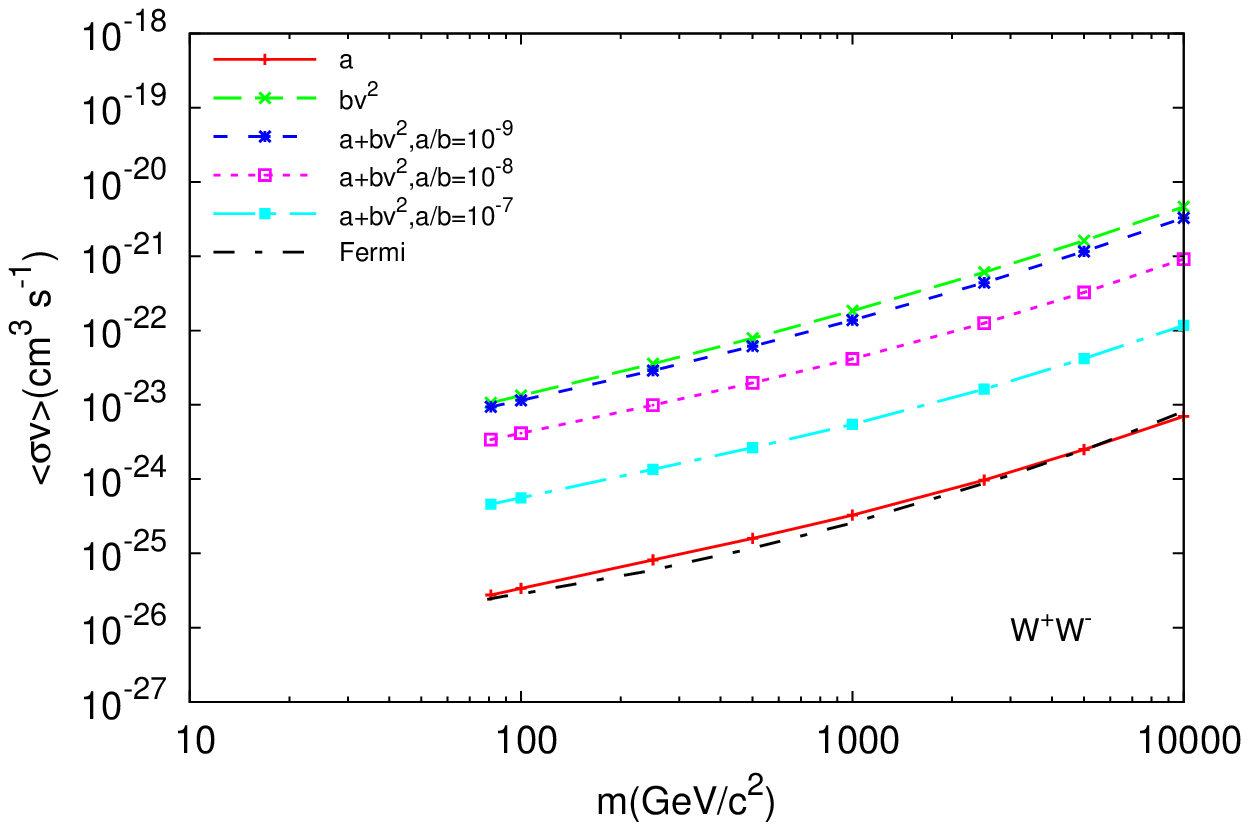}
\caption{Constraints on the DM annihilation cross section at $95\%$ C.L. for six annihilation channels from the six-year Fermi-LAT observation of six dSph's. The form of the DM annihilation cross section is assumed to be $a+b v^2$. The limits for five values of $a/b$ are shown. The dash dotted lines represent the limits set by the Fermi Collaboration \cite{2015arXiv150302641F}.
The contours for DM annihilations to $\mu^+\mu^-$ and $\tau^+\tau^-$
denote the parameter regions as the explanations of the anomalous cosmic-ray electrons and positrons observed by the AMS02 \cite{2015PhRvD..91f3508L}. "DR" and "DC" denote the diffusion-reacceleration and diffusion-convection cosmic-ray propagation models, respectively.}
\label{fig:zy2}
\end{figure*}

For a given ratio of $a/b$, the dSph $\gamma$-ray constrains on the parameters $a$ and $b$ can be obtained through the likelihood map method mentioned above. In order to make a comparison to the results from other DM detection experiments, these limits can be translated into the limits on the DM annihilation cross section in other astrophysical sources. For example, we can derive the limits on the local annihilation cross section with a velocity dispersion of $\sim 270$km s$^{-1}$. In Fig. \ref{fig:zy2}, we show these limits for several different values of $a/b$ for six DM annihilation channels.

Note that the cases for $a=0$ and $b=0$ denote the pure p-wave and s-wave annihilation, respectively. Our limits for the s-wave annihilation process have good agreement with those given by the Fermi-LAT Collaboration \cite{2015arXiv150302641F}. For a pure p-wave annihilation process, the constraints on the local
$\left<\sigma v\right>$ can be loosen by 3 orders of magnitude.

\subsection{Breit-Wigner scenario}

In the Breit-Wigner scenario, the initial DM particles annihilates via a pole which lies near twice the DM mass \cite{2009PhRvD..79f3509F,2009PhRvD..79i5009I,2009PhRvD..79e5012G,2009PhLB..678..168B}.
The mass of the resonance particle $M$ can be parametrized by
\begin{eqnarray}\label{eq:zy10}
M=2m_{\chi}\sqrt{1-\delta},
\end{eqnarray}
where $m_{\chi}$ is the DM mass, and $\delta$ is a parameter satisfying $|\delta|\ll1$.
A typical form of the DM annihilation cross section in the Breit-Wigner scenario can be written as
\begin{eqnarray}\label{eq:zy12}
\sigma v_{rel}=\frac{a}{(\delta+\frac{1}{4}v_{rel}^2)^2+\gamma^2},
\label{BWCS}
\end{eqnarray}
where $a$ is an undetermined parameter in the theoretical model. $\gamma$ is defined as
\begin{eqnarray}\label{eq:zy11}
\gamma=\Gamma/M,
\end{eqnarray}
where $\Gamma$ is the resonance decay width.

\begin{figure*}[!htb]
\centering
\includegraphics[width=0.9\columnwidth, angle=0]{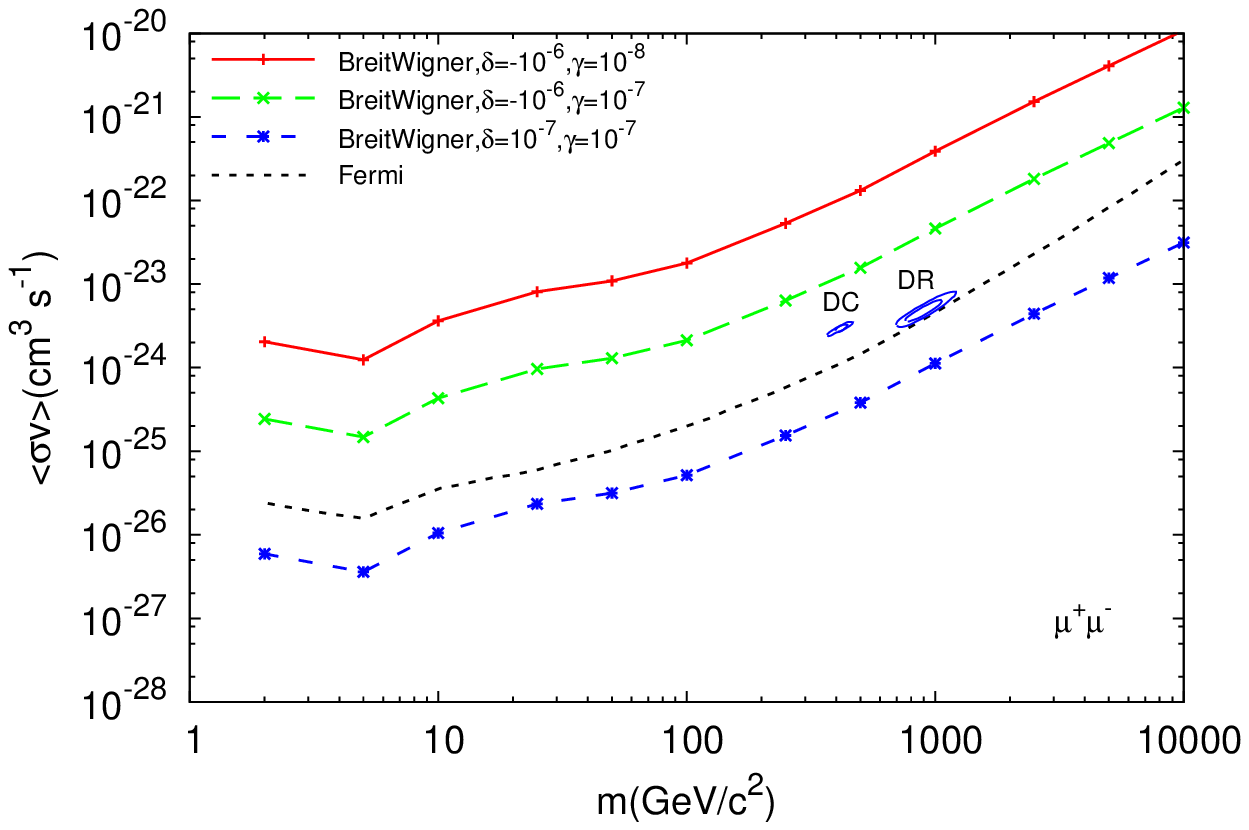}
\includegraphics[width=0.9\columnwidth, angle=0]{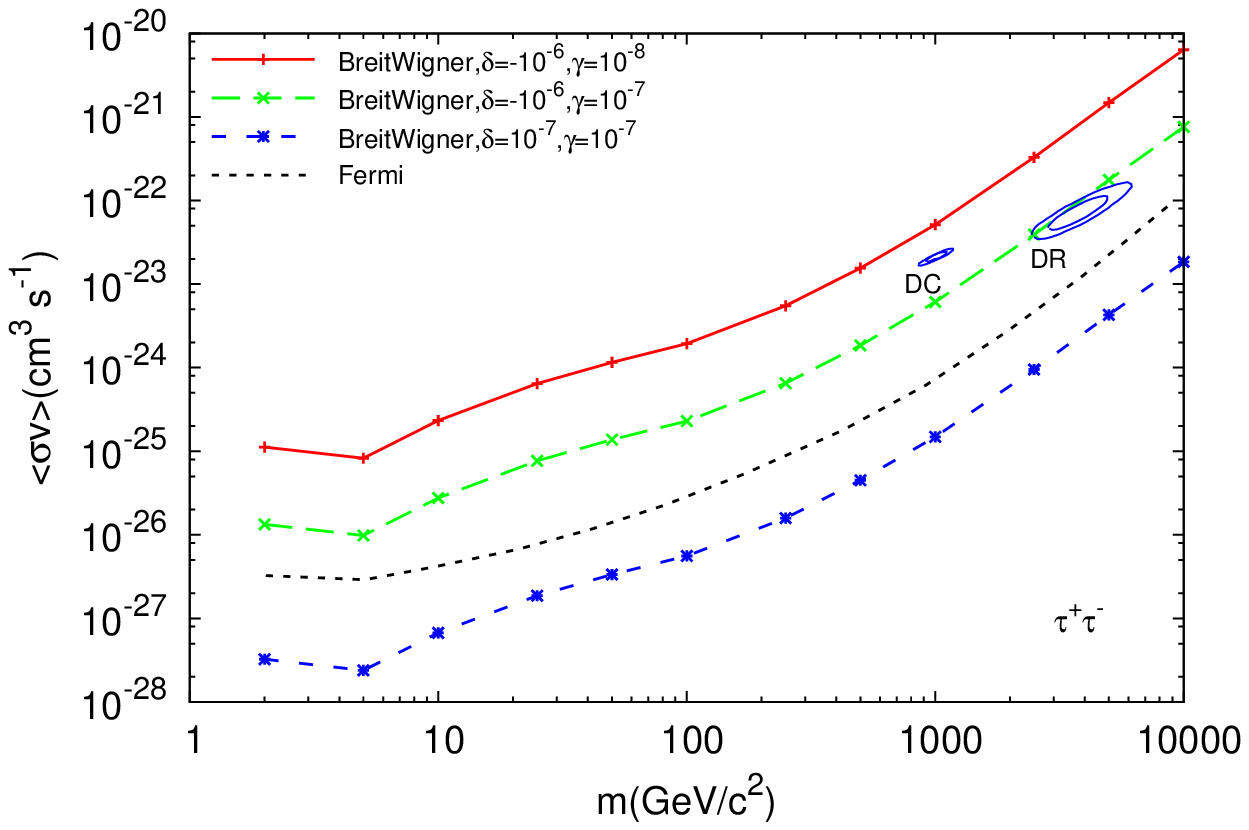}
\includegraphics[width=0.9\columnwidth, angle=0]{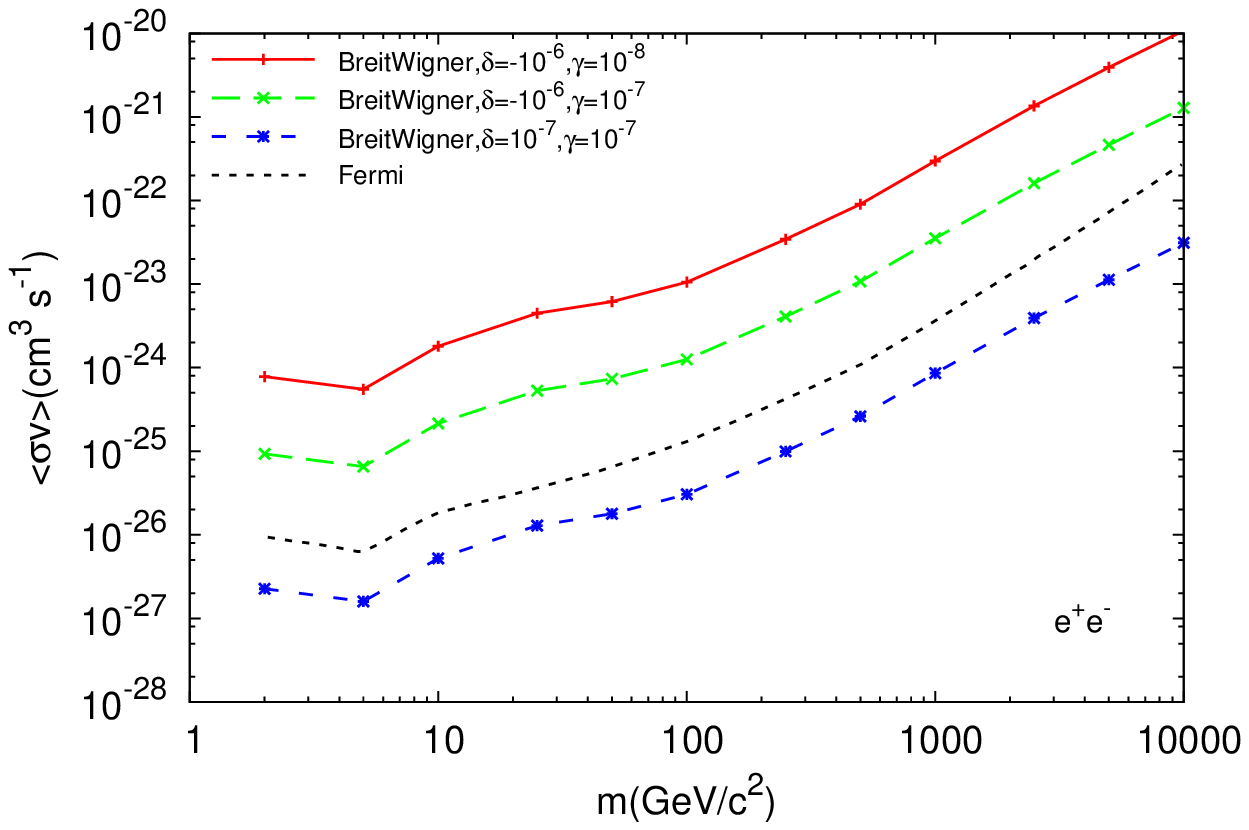}
\includegraphics[width=0.9\columnwidth, angle=0]{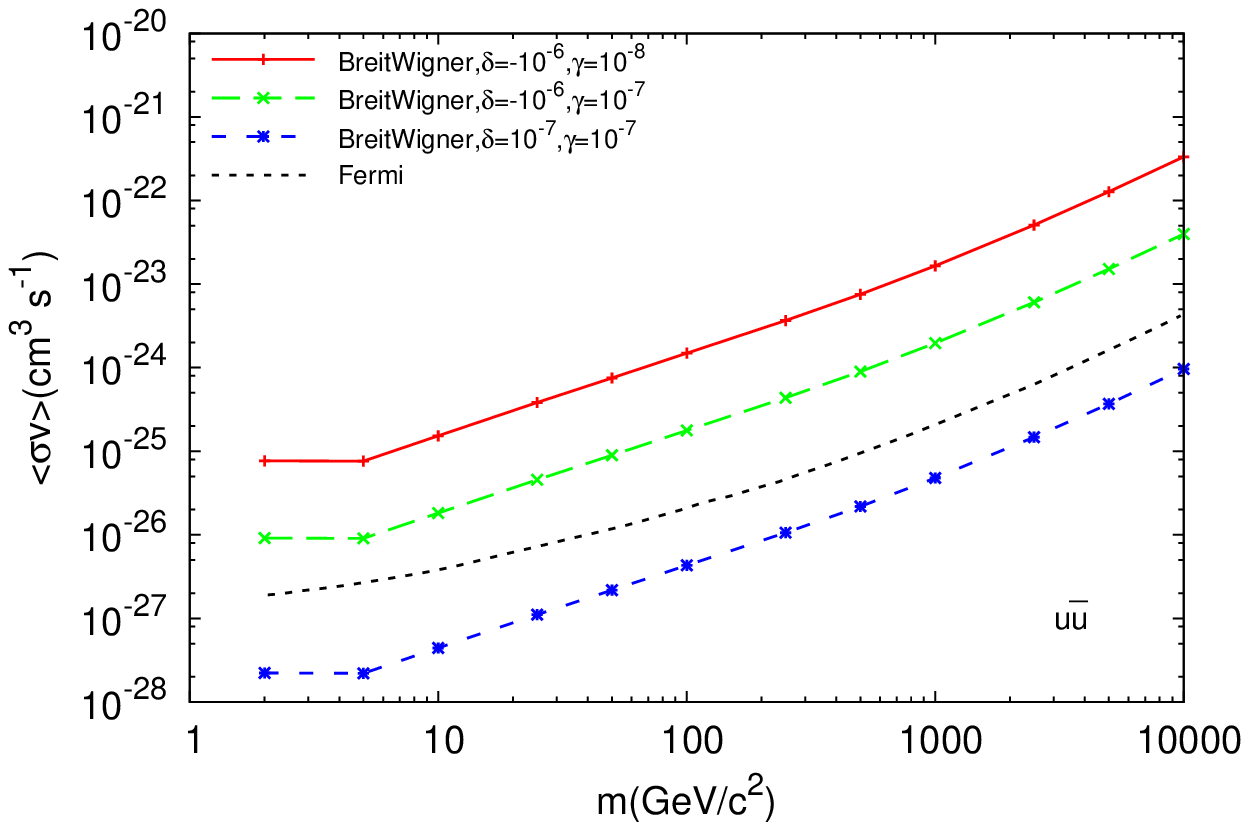}
\includegraphics[width=0.9\columnwidth, angle=0]{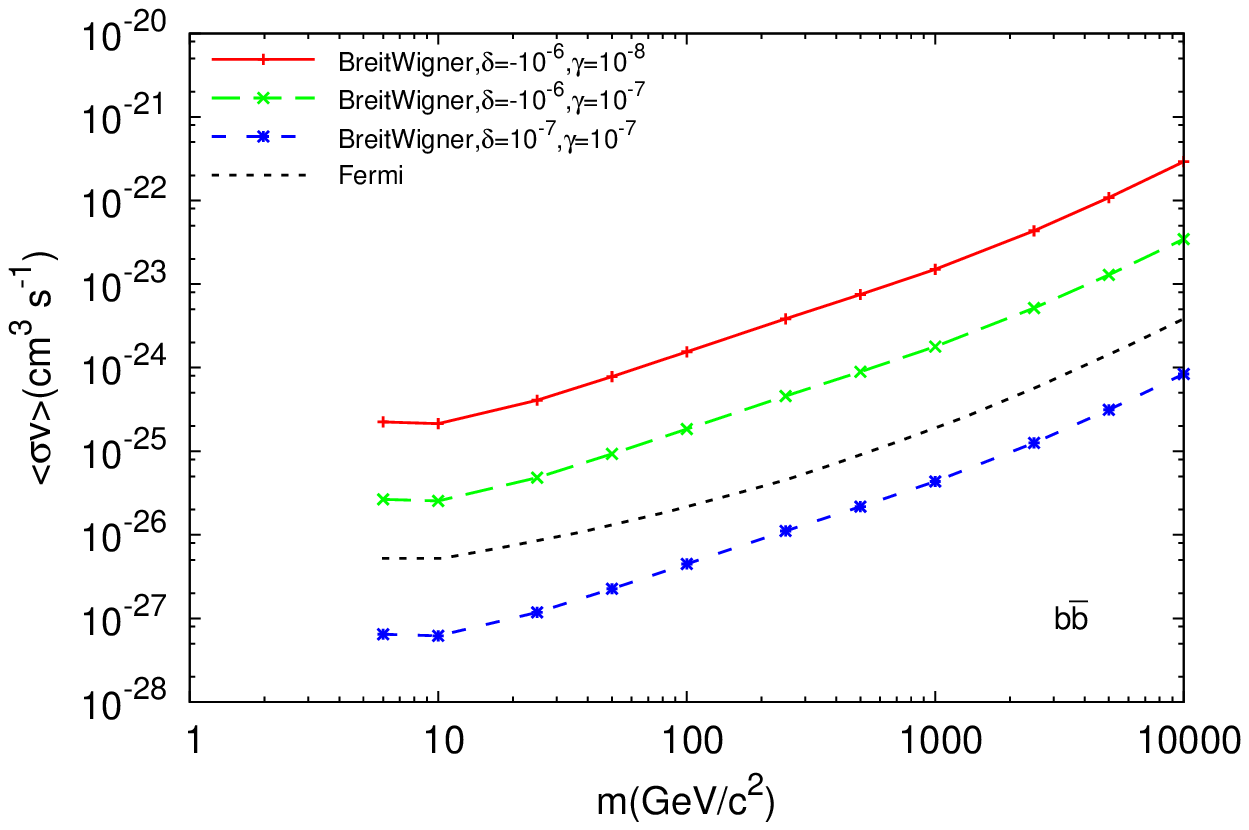}
\includegraphics[width=0.9\columnwidth, angle=0]{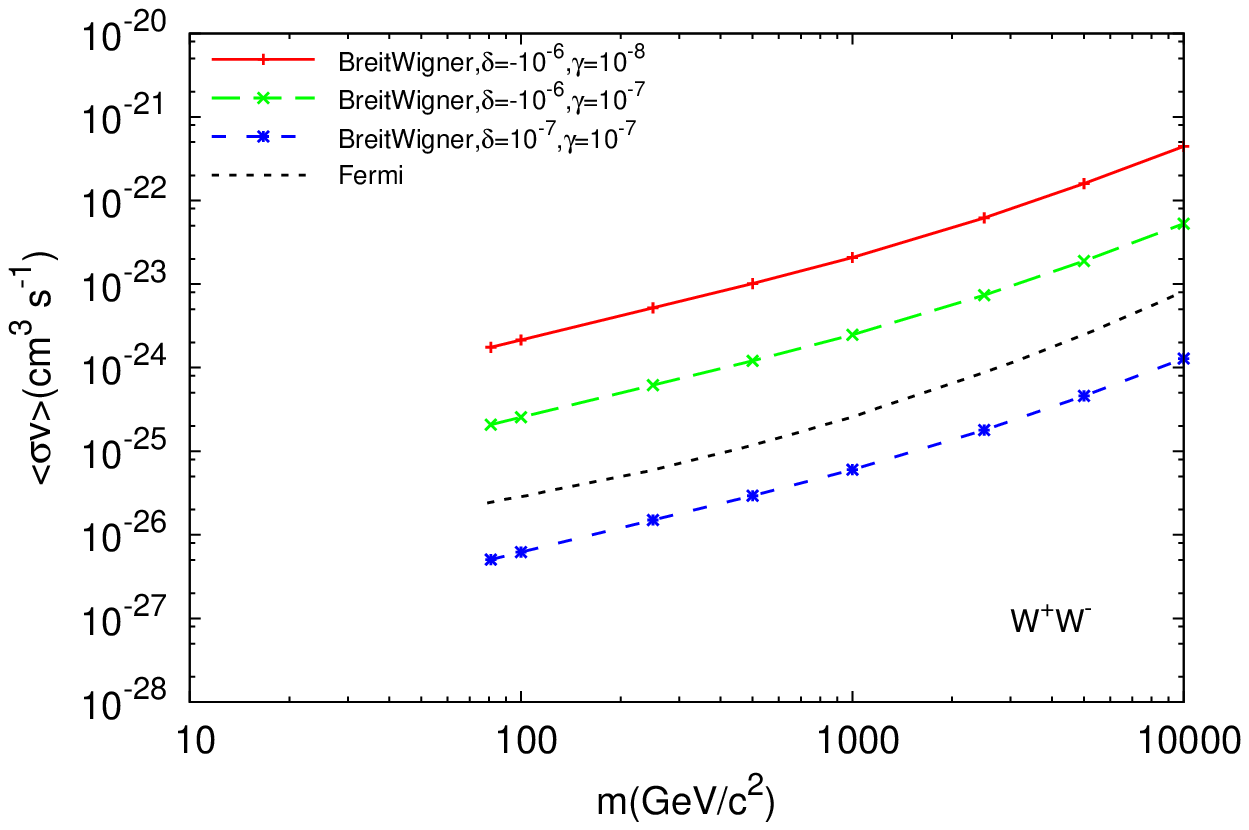}
\caption{Constraints on the DM annihilation cross section at $95\%$ C.L. for six annihilation channels from the six-year Fermi-LAT observation of six dSph's. The form of the DM annihilation cross section is taken as Eq. \ref{BWCS}. The limits for three sets of ${\delta, \gamma}$ are shown. The dash dotted lines represent the limits set by the Fermi Collaboration \cite{2015arXiv150302641F}.
The contours for DM annihilations to $\mu^+\mu^-$ and $\tau^+\tau^-$
denote the parameter regions as the explanations of the anomalous cosmic-ray electrons and positrons observed by the AMS02 \cite{2015PhRvD..91f3508L}. DR and DC denote the diffusion-reacceleration and diffusion-convection cosmic-ray propagation models, respectively.}
\label{fig:zy3}
\end{figure*}

For a given $\gamma$ and $\delta$, we can set constraints on $a$ from the Fermi-LAT dSph searches. The limits on the local DM annihilation
cross section for three sets of $\delta$ and $\gamma$ have been shown in Fig. \ref{fig:zy3}. For a comparison, we also show the parameter regions \cite{2015PhRvD..91f3508L} to explain the anomalous cosmic-ray electrons and positrons observed by the AMS02. For the case of $\delta>0$, the annihilation
cross section at lower velocities in the dSph's would always be larger than that at larger velocities in the local Galaxy. Thus, the dSph searches set very stringent limits on the parameter regions for the electron and positron excess. However, the annihilation cross section would be maximal at a velocity of $v\sim \sqrt{|\delta|}$ for $\delta<0$, and could be smaller at lower velocities in the range of $v < \sqrt{|\delta|}$. This means the limits from the dSph searches can be weaken if the dSph DM annihilations occur below the pole. This is the case for the limits of $\delta<0$ shown in
Fig. \ref{fig:zy3}.

\subsection{Sommerfeld scenario}

In this subsection, we consider the velocity dependent annihilation cross section scaled by a factor of $1/v$ or $1/v^2$  which can be obtained in the Sommerfeld scenario \cite{2005PhRvD..71f3528H,2007PhLB..646...34H,
2007NuPhB.787..152C,2009NuPhB.813....1C,2009PhRvD..79a5014A,2009PhLB..671..391P,2009PhRvD..79h3523L,2010PhLB..687..275D,
2010PhRvD..82h3525F,2010JPhG...37j5009C,2010JCAP...02..028S}. In this scenario, the long-range attractive interaction between two initial heavy DM particles via exchanges of light bosons $\phi$ can significantly enhance the annihilation cross section at low velocities, which can be expressed as
\begin{eqnarray}\label{eq:zy13}
\sigma v_{rel}
&\sim&(\sigma v_{rel})_0\frac{\pi\alpha_\chi/v}{1-e^{-\pi\alpha_\chi/v}} \nonumber\\
&\stackrel{\alpha_\chi \gg v}{\longrightarrow}& (\sigma v_{rel})_0\frac{\pi\alpha_\chi}{v} \nonumber\\
&\equiv&\frac{a}{v_{rel}},
\end{eqnarray}
where $(\sigma v_{rel})_0$ is the annihilation cross section in the early Universe, $\alpha_\chi$ is the interaction coefficient, and $a$ is a constant.
Moreover, in some special parameter regions, two incoming DM particles behave like a bound state; the annihilation cross section would be ``resonantly" (see e.g. \cite{2009PhRvD..79a5014A})
enhanced at low velocities as
\begin{eqnarray}\label{eq:zy14}
\sigma v_{rel} \sim (\sigma v_{rel})_0 \frac{\pi^2\alpha_{\chi}m_\phi}{6m_{\chi}v^2}.
\end{eqnarray}

In general, the Sommerfeld enhancement factor depends on several parameters, i.e. $m_\chi$, $m_\phi$, and $\alpha_\chi$, and can be obtained through solving the Schr\"{o}dinger equation with an attractive potential. We do not discuss this enhancement factor in a particular model. Instead, we only take two typical annihilation cross section forms which capture the key features:
\begin{eqnarray}\label{eq:zy15}
\sigma v_{rel}=\frac{a}{v_{rel}} \;\;  \mathrm{and} \;\;  \sigma v_{rel}=\frac{a}{v^2_{rel}}.
\label{SommCS}
\end{eqnarray}

\begin{figure*}[!htb]
\centering
\includegraphics[width=0.9\columnwidth, angle=0]{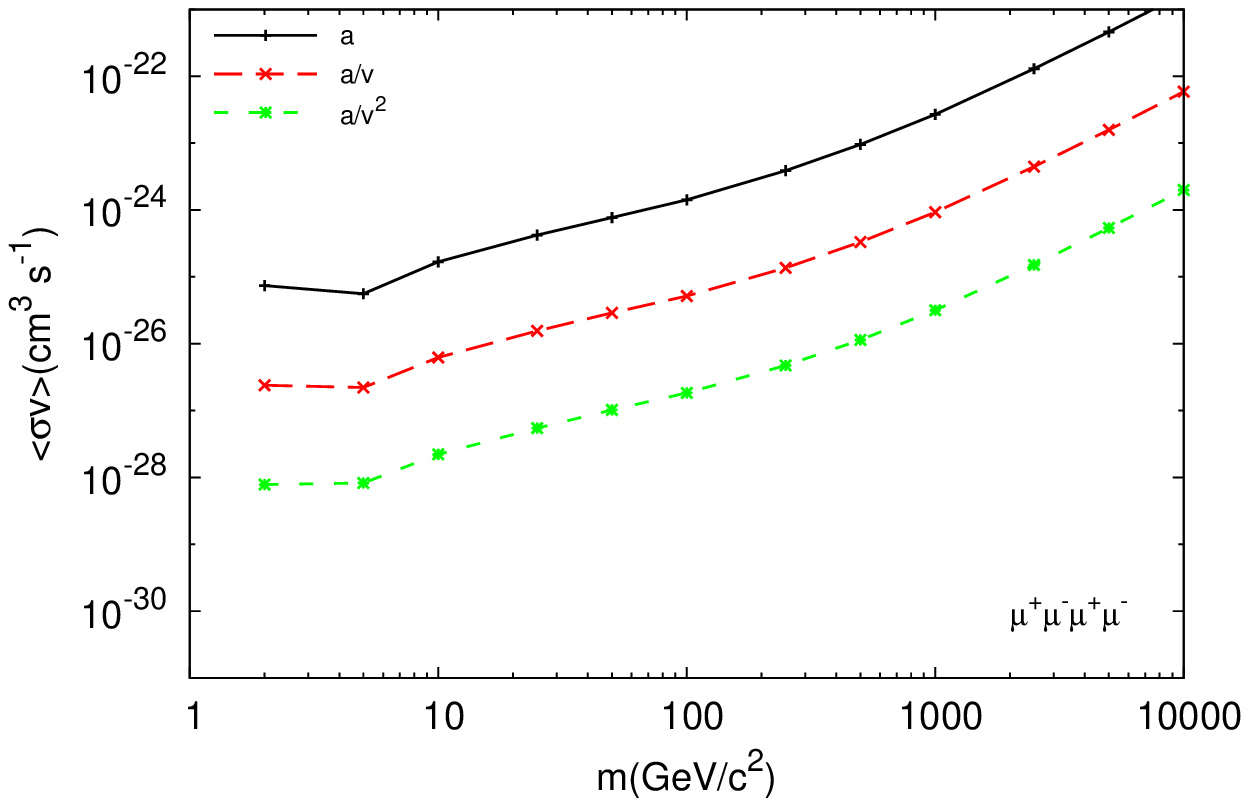}
\includegraphics[width=0.9\columnwidth, angle=0]{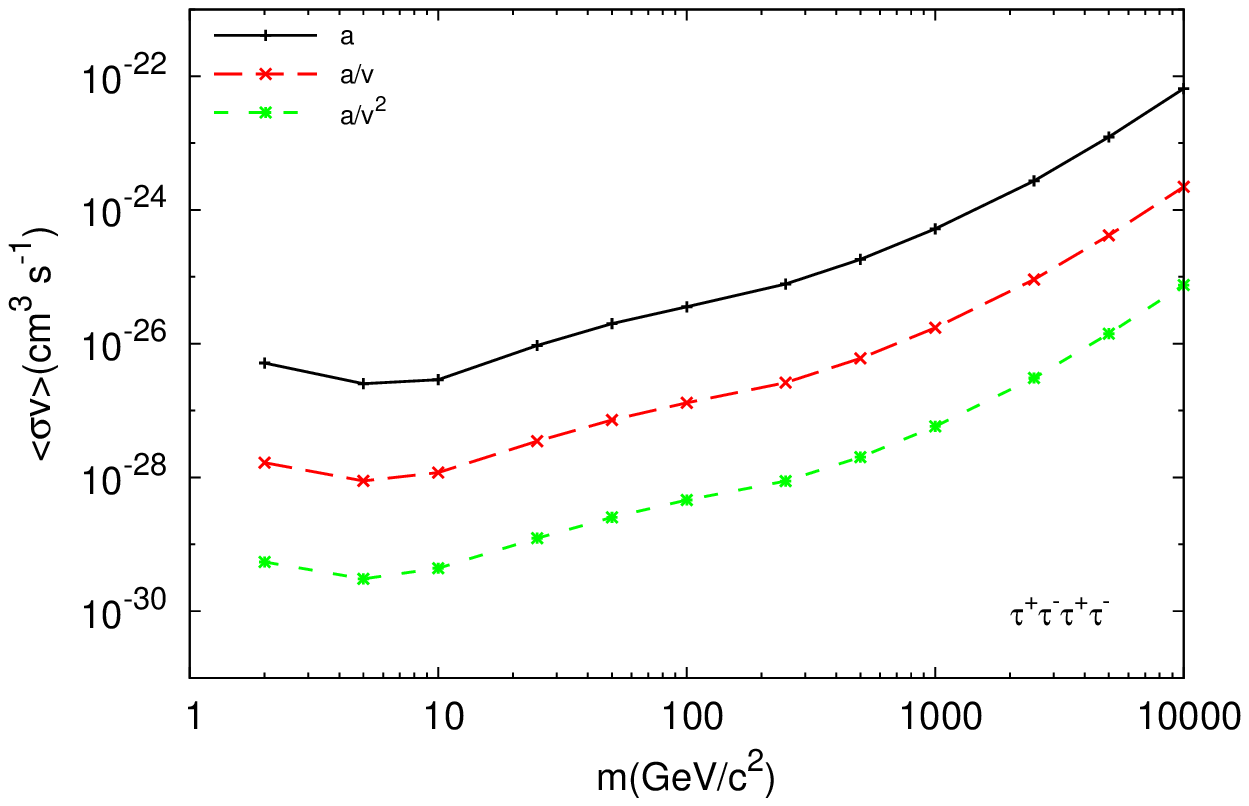}
\caption{Constraints on the DM annihilation cross section at $95\%$ C.L. for two annihilation channels from the six-year Fermi-LAT observation of six dSph's. The DM annihilation cross section is assumed to be proportional to a factor of $1/v$ or $1/v^2$. The black solid lines represent the limits on the velocity dependent annihilation cross section.}
\label{fig:zy4}
\end{figure*}

The limits on the local DM annihilation cross section with the form of Eq. \ref{SommCS} for two annihilation channels $\chi \chi \rightarrow 4\mu$ and $4\tau$ are shown in Fig \ref{fig:zy4}. For a comparison, we also give the limits for the velocity independent annihilation. In general, the four body annihilation final states would induce weaker limits because of the softer initial $\gamma-$ray spectra. However, since the DM annihilation cross sections are enhanced at low velocities in this scenario, the dSph observations would set much stricter limits than those from the galactic observations.

\section{Conclusion and discussions}

In this work, we study the limits on the velocity dependent DM annihilation cross section from the Fermi-LAT dSph $\gamma-$ray observations
based on the latest Pass8 data. We construct the likelihood maps in the energy range of $0.5-500$ GeV for six nearby luminous dSph's with large $J$ factors. Then we can easily obtain the total likelihood, and derive the limit on the DM annihilation cross section for an arbitrary initial $\gamma$ ray spectrum.

In the analysis, we consider three typical velocity
dependent annihilation cross section forms. Since the DM particles in different astrophysical sources often have different
velocity dispersions, the annihilation rates may dramatically change. For DM annihilation with a non-negligible p-wave contribution, the
cross section can be parametrized as $a+bv^2$, and is suppressed in the dSph's because of their low velocity dispersions.
For the pure p-wave annihilation, the limits can be weaken by about 3 orders of magnitude more than those for the s-wave annihilation. For the annihilation cross section scaled by a factor of $1/v$ or $1/v^2$, which can be obtained in the Sommerfeld scenario, the dSph constraints would be stronger. For other more complex velocity dependent annihilation cross section forms, the behavior of the constraints depends on the model parameters.
As an example in the Breit-Weigner scenario with the particular parameter sets, the dSph constraints may be weaker or stronger than those for the velocity independent annihilation.

In some cases, the limits on the local DM annihilation cross section by
the Fermi-LAT observation of dSph's can be relaxed.
An important implication of such a fact is the DM explanation of the cosmic-ray positron/electron
excess observed by AMS-02.
Contrary to the velocity independent annihilation scenario, where the DM explanation has almost been excluded by the Fermi-LAT data, DM annihilation in
velocity dependent scenarios may remain viable
to explain the positron/electron excess.

\acknowledgments{This work is supported by the National Natural Science Foundation of China under Grants Nos.~11475189, 11475191, 11135009, 11175147, the 973 Program of China under
Grant No.~2013CB837000, and
by the Strategic Priority Research Program
``The Emergence of Cosmological Structures'' of the Chinese
Academy of Sciences under Grant No.~XDB09000000. We are grateful to Q. Yuan, X.-Y. Huang, Z.-W. Li, S.-J. Lin and Q.-F. Xiang for helpful discussions.}

\end{document}